\documentclass[12pt]{article}
\usepackage{amsmath,amsfonts,amsthm,amssymb}
\numberwithin{equation}{section}
\usepackage{setspace}
\usepackage{graphicx}
\usepackage{cancel}

\newcommand{\Tr}{{\rm Tr\,}}
\newcommand{\nn}{\nonumber}
\newcommand{\SL}{{\rm SL}}
\newcommand{\sla}{{\mathfrak{sl}}}
\newcommand{\ol}[1]{{\overline{#1}}}
\newcommand{\AdS}{{\rm AdS}}
\newcommand{\dS}{{\rm dS}}

\begin{document}

\begin{titlepage}
\hfill MCTP-13-25

\begin{center}
{\Large \bf  On Resolutions of Cosmological Singularities}

\vskip .7 cm

{\Large \bf  in Higher-Spin Gravity}\\
\end{center}

\vskip .7 cm

\vskip 1 cm
\begin{center}
{ \large Benjamin Burrington${}^a$, Leopoldo A. Pando Zayas${}^b$ and Nicholas Rombes${}^{b}$}
\end{center}

\vskip .4cm
\centerline{\em ${}^a$ Department of Chemistry and Physics, Troy University}
\centerline{\it  Troy, AL 36082}

\bigskip\bigskip

\centerline{\it ${}^b$ Michigan Center for Theoretical
Physics}
\centerline{ \it Randall Laboratory of Physics, The University of
Michigan}
\centerline{\it Ann Arbor, MI 48109-1120}

\bigskip\bigskip

\vskip 1 cm

\vskip 1.5 cm
\begin{abstract}
We study the resolution of certain cosmological singularity in the context of higher-spin three-dimensional gravity.  We consider gravity coupled to a spin-3 field realized as Chern-Simons theory with gauge group $SL(3,\mathbb{C})$. In this context we elaborate and extend a singularity resolution scheme proposed by Krishnan and Roy. We discuss the resolution of a big-bang singularity in the case of gravity coupled to a spin-4 field realized as Chern-Simons theory with gauge group $SL(4,\mathbb{C})$. In all these cases we show the existence of gauge transformations that do not change the holonomy of the Chern-Simons gauge potential and lead to metrics without the initial singularity. We argue that such transformations always exist in the context of gravity coupled to a spin-N field when described by Chern-Simons with gauge group  $SL(N,\mathbb{C})$.
\end{abstract}

\end{titlepage}

\tableofcontents

\section{Introduction}

Einstein theory of general relativity has passed many stringent experimental tests and provides the pillars to our current understanding of cosmology and black hole physics. It is not, however, compatible with quantum mechanics and one therefore expects it to be subsumed into some other theory, such as string theory. It is thus natural to consider modifications to Einstein's equations. One such modification that has attracted a lot of attention recently is higher-spin gravity in asymptotically AdS spacetimes as constructed by Vasiliev \cite{Vasiliev:1999ba}\cite{Bekaert:2005vh}. This extension of the Einstein's equations of motion is very constrained and has a number of distinctive properties. Higher spin theories are predominantly defined as classical theories. In some cases, however, it is argued that higher spin theories are a ``corner'' of string theory. Namely, some higher-spin theories are conjectured to be related to the tensionless limit of string theory \cite{Sezgin:2002rt,HaggiMani:2000ru,Beisert:2004di}. This limit can also be understood as the ultra-high energy limit of string theory \cite{Gross:1987ar, Gross:1988ue} where stringy effects are dominant.

In this paper we will focus on the three-dimensional version of higher spin gravity. Three-dimensional higher spin theories are simpler to work with than the general case for several reasons. First, they admit a truncation whereby spin less than a certain number $N$ can be consistently retained  \cite{Blencowe:1988gj,Bordemann:1989zi,Henneaux:2010xg,Campoleoni:2010zq}; in higher dimensions it is inconsistent to retain a finite number of higher-spin fields \cite{Vasiliev:1990en}. Another important feature of three-dimensional higher-spin theories is that, rather than a complicated set of equations of motion as is the case in higher dimension, they admit a relatively simple description in terms of Chern-Simon theories with gauge group $SL(N,\mathbb{R})\times SL(N, \mathbb{R})$ or $SL(N,\mathbb{C})$, \cite{Henneaux:2010xg}, \cite{Campoleoni:2010zq} \cite{Campoleoni:2012hp} \cite{Banados:2012ue}.  Higher spin theories also fit naturally in the general context of the AdS/CFT correspondence. Indeed, many interesting aspects of the $AdS_3/CFT_2$ holography with higher-spin fields have been developed recently, see for example: \cite{Gaberdiel:2010pz,Gaberdiel:2011wb,Gaberdiel:2011zw,
Creutzig:2011fe,Chang:2011vka,Candu:2012jq,Gaberdiel:2012ku,
Gaberdiel:2012uj,Candu:2012tr,Creutzig:2012ar,Candu:2012ne,Peng:2012ae,
Chang:2013izp}.

Any theory of gravity has to provide answers to questions raised in the context of black hole physics and in the cosmological setup. Indeed, various interesting aspects of black hole physics have been discussed in the context of three-dimensional higher-spin gravity by considering modifications of the BTZ black hole \cite{Banados:1992wn,Banados:1992gq}. For example, questions such as explicit constructions  of black holes \cite{Gutperle:2011kf},\cite{Ammon:2011nk}, \cite{Kraus:2011ds} see in particular the review \cite{Ammon:2012wc}; definition of thermodynamics rules \cite{Perez:2012cf,David:2012iu,Perez:2013xi,deBoer:2013gz,Kraus:2013esi}. Glaringly, a gauge-invariant definition of horizon is lacking but a lot of effort is going into the construction of a coherent picture by using various techniques  \cite{Ammon:2013hba,deBoer:2013vca},  interesting questions about unitarity and singularity resolution have also been addressed \cite{Castro:2012bc,Castro:2011fm}.

The cosmological aspects of higher-spin gravity have received less attention. One interesting effort in this direction concerns the status of cosmological singularities in higher spin \cite{Krishnan:2013cra}. One goal of this paper is to discuss the question of cosmological singularities in general and big-bang singularity in particular in the context of higher-spin gravity. We focus on a precise formulation of the problem and do not address potentially related questions such as the dS/CFT correspondence, see \cite{Ouyang:2011fs,deBuyl:2013ega} for a discussion of those topics. There is already a very rich history of resolutions of cosmological singularities in string theory, see for example: \cite{Cornalba:2002fi,Berkooz:2002je,Johnson:2004zq,Balasubramanian:2002ry} and the review \cite{Cornalba:2003kd} for more references; more recent works include also \cite{Craps:2007ch,Biswas:2011ar,Bagchi:2012xr}. We hope that higher-spin gravity would provide another interesting playground to discuss the resolution of cosmological singularities.

The paper is organized as follows. In section \ref{Review} we review the Chern-Simons formulation of three-dimensional gravity and the work \cite{Krishnan:2013cra}; we also point out fundamental differences between the situation for spin-3 black holes and spin-3 cosmology. In section \ref{Review}, after reviewing a resolution proposed by Krishnan and Roy in \cite{Krishnan:2013cra} we also show a more general approach to the resolution of singularities in spin-3 gravity. Section \ref{Spin4} describes the resolution of the big-bang singularity in the context of gravity coupled to a spin-4 field, we consider two cases with varying degree of generality. In section \ref{SpinN} we show how the resolution of singularities is possible in the context of spin-N gravity for arbitrary $N$. We conclude in section \ref{Conclusions}. We relegate some technical aspects and brief discussions of geometrical aspects as viewed from the diffeomorphic-invariant point of view to the appendices.


\section{Three-dimensional higher spin gravity as a Chern-Simons theory}\label{Review}
Let us first review how in three dimensions Einstein gravity can be written in terms of Chern-Simons theory \cite{Witten:1988hc} \cite{Achucarro:1987vz}. In particular, we shall review how dS${}_3$ appears as a solution of Chern-Simons with $SL(2,\mathbb{C})$ gauge group. On the Chern-Simons side the starting point is a pair of connections taking values in the algebra $\sla(2)$ with generators $\{L_0, L_1, L_{-1}\}$ (see appendix (\ref{Generators}) for an explicit realization of the generators).

The Chern-Simons connections encode the vielbein, $e$, and the spin connection, $\omega$, as
\begin{equation} \label{eq:csConnection}
  A=\omega+ie, \hspace{1cm} \ol{A}=\omega-ie,
\end{equation}
from these data the metric can be found as
\begin{equation} \label{eq:csMetric}
  g_{\mu\nu}=\frac{1}{\Tr(L_0L_0)}\Tr(e_\mu e_\nu).
\end{equation}
Other important aspects beyond these classical identifications and pertaining to the quantum theory were first discussed in \cite{Witten:1988hc}.

\subsection{Review of  spin-3 cosmology and singularity resolution}
In this section we review the construction of \cite{Krishnan:2013cra} and set up the notation for our generalization.
The starting metric, written in Fefferman-Graham form, is
\begin{equation} \label{eq:dsMetric}
  ds^2=\frac{2\pi}{k}(\mathcal{L}(x)dx^2+\ol{\mathcal{L}}(\ol{x})d\ol{x}^2)+\left(e^{2\tau}
  +\left(\frac{2\pi}{k}\right)^2\mathcal{L}(x)\ol{\mathcal{L}}(\ol{x})e^{-2\tau}\right)dxd\ol{x}-d\tau^2.
\end{equation}
Note that for $\mathcal{L}=\ol{\mathcal{L}}=0$  the above metric becomes $ds^2=-d\tau^2 +e^{2\tau}dx d\ol{x}$, which corresponds to dS${}_3$ (in planar coordinates) with unit radius. To connect with more traditional coordinates, it is convenient to consider $x=\phi+iz$, $\ol{x}=\phi-iz$. The planar de Sitter metric plays and important role in cosmological models. It describes a spatially flat universe undergoing exponential expansion. Such models are a good description of either the inflationary epoch or the present epoch where the evolution of the universe is dominated by a positive cosmological constant \cite{DInverno}. The planar dS solution contains a big bang singularity at $\tau \to -\infty$ and causal horizon, see \cite{HawkingEllis} and \cite{Spradlin:2001pw} for a complete description of this classical properties including the full Penrose diagram.

In the case of $\mathcal{L}$ and $\ol{\mathcal{L}}$ nonvanishing the structure is more intricate. We will be particularly interested in the possibility of removing a singularity which can be seen in the volume of the space:
\begin{equation}
  \sqrt{\det g}=\frac{1}{2}e^{2\tau}-\frac{2\pi^2}{k^2}\mathcal{L}\ol{\mathcal{L}}e^{-2\tau}.
\end{equation}
Note that there is singularity at $e^{4\tau}=\frac{4\pi^2}{k^2}\mathcal{L}\ol{\mathcal{L}}$, at which the space has vanishing volume.

When considered in the more general context of spin-3 gravity we would like to show that the aforementioned singularity is an artifact of a gauge choice, that is, that there exist gauge transformations in $\sla(3)$ that lead to metrics without singularities. This is a  manifestation of the fact that spacetime (diffeomorphism-invariant information from the spin-2 field) is not generally invariant under higher-spin gauge transformations valued in the higher-spin group. \\

One can check that the following connection reproduces the metric (\ref{eq:dsMetric}) via the definitions \eqref{eq:csConnection} and \eqref{eq:csMetric}:
\begin{equation} \label{eq:spin2connectionA}
\begin{split}
  A&=\left(e^\tau L_1+\frac{2\pi}{k}e^{-\tau} \mathcal{L}(x) L_{-1}\right)dx+ L_0d\tau, \\
  \ol{A}&=-\left(e^\tau L_{-1}+\frac{2\pi}{k}e^{-\tau} \ol{\mathcal{L}}(\ol{x}) L_{1}\right)d\ol{x}-L_0d\tau.
\end{split}
\end{equation}
To resolve the singularity, the goal is to find a new connection $A'$ that has the same holonomy as $A$ (so that they are related by a single-valued gauge transformation) and modifies the metric so as to remove the singularity. The generalization considered in \cite{Krishnan:2013cra} is as follows:
\begin{equation}
\begin{split}
  A'&=A+Xdx,\hspace{1cm} X\equiv\sum_{a=-2}^2 e^{a\tau}C_a(\tau)W_a, \\
  \ol{A'}&=\ol{A}+\ol{X}d\ol{x},\hspace{1cm}\ol{X}\equiv\sum_{a=-2}^2 e^{a\tau}\ol{C}_a(\tau)W_a,
\end{split}
\end{equation}
where the coefficients $C_a(\tau)$ are not allowed to depend on $x$ or $\ol{x}$, in order to simplify the calculation of the holonomy matrix later.

Chern-Simons theory is a theory of flat connections as follows from the equations of motion which are
\begin{equation}
  dA'+A'\wedge A'=0.
\end{equation}
In order to simplify these equations, we can move to a primitive, $\tau$-independent connection $a'$, which is related to $A'$ by a single-valued gauge transformation:
\begin{equation} \label{eq:gaugeTransform}
  A'=b^{-1}a'b+b^{-1}db, \hspace{1cm} b=e^{\tau L_0}.
\end{equation}
Then the equations of motion
\begin{equation}
  da'+a'\wedge a'=0,
\end{equation}
become $\partial_\tau C_a=0$; thus $C_a$ must be constant.

It turns out that the following conditions on our constants are sufficient to ensure singularity resolution:
\begin{equation} \label{eq:revConds}
  \Tr(X^2)=\Tr(\ol{X}^2)=0,\hspace{1cm}\Tr(X\ol{X})>0.
\end{equation}
These explicitly become
\begin{equation}
\begin{split}
  0&=C_0^2-3C_1C_{-1}+12C_2C_{-2}, \\
  0&<C_0\ol{C}_0-3C_1\ol{C}_{-1}+12C_2\ol{C}_{-2}+\text{c.c.}
\end{split}
\end{equation}
The final condition is that $A'$ and $A$ must have the same holonomy. Instead of working directly with the eigenvalues of the holonomy matrix, we will enforce the equivalence of the coefficients in the characteristic polynomials of each matrix. The characteristic polynomial of a general $3\times 3$ matrix $M$ can be written as
\begin{equation}
  p_M(\lambda)=\det(M)-\frac{1}{2}(\Tr(M)^2-\Tr(M^2))\lambda+\Tr(M)\lambda^2-\lambda^3.
\end{equation}
Since the holonomy matrices reside in $\sla(3)$, they are traceless. The role of the characteristic polynomial in the context of higher spin  black holes was highlighted in \cite{Gutperle:2011kf}. Thus, the two conditions we impose are
\begin{equation}
  \det(H[A'])=\det(H[A]),\hspace{1cm}\Tr(H[A']^2)=\Tr(H[A]^2),
\end{equation}
where the holonomy matrix of a connection $A$ is
\begin{equation}
  H[A]=\int A|_{z=const,\tau=0}=\int_0^{2\pi}d\phi\,A_{\phi}|_{\tau=0}.
\end{equation}
The second condition above is equivalent to the condition we already found, so there is only one new constraint, in the form of a complex equation. The specific form of this new constraint is dependent upon the choice of $\sla(3,\mathbb{C})$ generators.

Summarizing, the original Ansatz had 10 degrees of freedom, in the form of the five complex constants $C_a$. We imposed two complex constraints and one inequality, so we are left with $10-4=6$ degrees of freedom parametrizing  the singularity-resolving gauge transformation.

\subsection{Higher spin cosmology and higher-spin black holes}\label{hsc-hsbh}

In this subsection we compare the situation in the cosmological setup with what is known in the context of higher-spin black holes. First, let us recall some of the standard facts that have been used in the context of constructing and analyzing higher-spin black holes \cite{Gutperle:2011kf,Ammon:2011nk,Ammon:2012wc}. This technology can be borrowed and applied to the cosmological context. In particular, we think it is important to identify what makes this singularity resolution possible, and what makes the cosmology case different from the black hole case. The key lies in the $\tau$-dependence of the conjugate connection, and how it changes the form of the veilbeins in each case.

Let us consider first the spin-3 black hole.  The connection $A$, which is expanded in a basis of $\sla(3)$ generators, has factors of $e^{kr}$ attached to each generator of weight $k$:
  \begin{equation}
    A=A^a T^a_{(k)}=\mathcal{A}^a T_{(k)}^a e^{kr}
  \end{equation}
  where $T_{(k)}^a$ is a generator of $\sla(3)$ of weight $k$ \cite{Gutperle:2011kf,Ammon:2011nk,Ammon:2012wc}. A similar pattern is followed in the construction of spin-4 black holes as presented in  \cite{Tan:2011tj} and \cite{Ferlaino:2013vga}.  The right connection, $\ol{A}$, has factors of $e^{-kr}$ attached to each generator of weight $k$:
    \begin{equation}
      \ol{A}=\ol{A}^a T^a_{(k)}=\ol{\mathcal{A}}^a T_{(k)}^a e^{-kr}
    \end{equation}
  Note that in this case, $\mathcal{A}^a$ and $\ol{\mathcal{A}}^a$ are \emph{not} complex conjugates. Thus, each Dreibein $e^a=\frac{1}{2}(A^a-\ol{A}^a)$ has a term proportional to $e^{kr}$ and a term proportional to $e^{-kr}$.

Traces of pairs of generators of $\sla(3)$ are nonzero only if the weights of the generators add up to zero. This is not merely a feature of the principal embedding of $\sla(2)$ into $\sla(4)$ which we are using, it is  a general feature of the algebra (see appendix (\ref{Generators}) for an explicit proof).

Thus,  when a field is constructed, e.g.,  the metric $g_{\mu\nu}=\frac{1}{\Tr(L_0L_0)}\Tr(e_\mu e_\nu)$, it will have terms proportional to $e^{2k r}$ and $e^{-2kr}$, where $k$ can be the weight of any generator. In particular, in the spin-3 case in literature, the black hole metric has terms proportional to $e^{4r}$; that is, extending the gauge group to $ SL(N,\mathbb{R})$ has the consequence of changing the asymptotic behavior of spacetime, so that the space is no longer asymptotically AdS of the same radius. A non-vanishing higher-spin charge is achieved only when a chemical potential for the corresponding higher-spin current is turned on. This chemical potential can be viewed as an irrelevant deformation (see \cite{Gutperle:2011kf,Ammon:2011nk,Ammon:2012wc}). Interestingly, although the UV asymptotic of the original geometry is changed, it can be interpreted as a flow to a new asymptotically AdS${}_3$ geometry with a different radius and appropriately different asymptotic symmetry algebra (see also \cite{Tan:2011tj} and \cite{Ferlaino:2013vga} for a discussion of spin-4 case.).

      Let us now review why the construction of cosmological solutions is different in structure. The connection $A$, which is expanded in a basis of $\sla(3)$ generators, has factors of $e^{k\tau}$ attached to each generator of weight $k$, just as in the black hole case. The crucial difference is that the conjugate connection $\ol{A}$ \emph{also} has factors of $e^{k\tau}$ attached to each generator of weight $k$, since if $A=A^a T^a$, the coefficients of the conjugate connection, $\ol{A}^a$, are simply the complex conjugates of $A^a$. Thus, each Dreilbein $e^a=\frac{1}{2i}(A^a-\ol{A}^a)$ has an \emph{overall} factor of $e^{k\tau}$:
    \begin{equation}
      e=e^aT_{(k)}^a=\frac{1}{2i}(\mathcal{A}^a-\ol{\mathcal{A}}^a)T^a_{(k)}e^{k\tau}
    \end{equation}

As in the black hole case, traces of pairs of generators of $\sla(4)$ are nonzero only if the weights of the generators add up to zero (see appendix (\ref{Generators}) for an explicit proof.).

Therefore,  when a field is constructed, i.e. the metric $g_{\mu\nu}=\frac{1}{\Tr(L_0L_0)}\Tr(e_\mu e_\nu)$, the $\tau$-dependence of the vielbeins will cancel out, since the sum of the weights is zero (except for the case of $L_{-1}$ and $L_1$, which is different because it represents the embedding of normal spin-2 gravity into the higher-spin theory):

    \begin{equation}
    \begin{split}
      g_{\mu\nu}&\propto \Tr(T^a_{(k)}T^b_{(k)})e^a_\mu e^b_\nu \\
      &\propto \Tr(T^a_{(k)}T^b_{(k')})(\mathcal{A}^a_\mu-\ol{\mathcal{A}}^a_\mu)(\mathcal{A}^b_\nu-\ol{\mathcal{A}}^b_\nu)e^{(k+k')\tau} \\
      &\propto \Tr(T^a_{(k)}T^b_{(-k)})(\mathcal{A}^a_\mu-\ol{\mathcal{A}}^a_\mu)(\mathcal{A}^b_\nu-\ol{\mathcal{A}}^b_\nu)+\text{terms from pure gravity}
    \end{split}
    \end{equation}
    In the last line, $a$ and $b$ do not run over the indices consisting of the pure gravity generators, which in our paper are $L_{-1}$, $L_0$, $L_1$---the terms relating to pure gravity give the metric its $\tau$-dependence. This preserves the cosmological structure for early and late times.
In addition to giving rise to a ``simpler'' metric, the $\tau$-dependence cancellation allows us to impose time-independent conditions for singularity resolution, in the form of \eqref{eq:revConds}. In the black hole case, these conditions would be $r$-dependent, but in this case, the cosmology case, the cancelation of $\tau$ allows us to add a constant positive-definite quantity to the volume.

\subsection{General connection and general singularity resolution}

Inspired by the analogy with the black hole context, we consider a more general connection  in this subsection. The goal is to show that the singularity can be resolved via a gauge transformation that of general type. Our initial Ansatz is reminiscent of the black hole Ansatz used in
\cite{Gutperle:2011kf,Ammon:2011nk,Ammon:2012wc}  and extended in \cite{Tan:2011tj} and \cite{Ferlaino:2013vga}.

We begin with an ansatz for an $\sla(3,\mathbb{C})$-valued connection (see appendix \ref{sl3genapx} for explicit generators) which is a generalization of \eqref{eq:spin2connectionA}, in parallel with the current literature for spin-4 black holes in $\AdS_3$ (note that this is the most general asymptotically $\dS_3$ connection):
\begin{equation}
\begin{split}
  a'&=\left(L_1+\mathcal{L}L_{-1}+\mathcal{W}W_{-2}\right)dx \\
  &\hspace{1cm}+(w_{-2}W_{-2}+w_{-1}W_{-1}+w_0W_0+w_1W_1+w_2W_2+l L_{-1})d\ol{x}
\end{split}
\end{equation}
where all of the coefficients of the generators belong to $\mathbb{C}$. Here we omit the normalization coefficients of $\mathcal{L}$ and $\mathcal{W}$ for simplicity. They can be restored by $\mathcal{L}\rightarrow -\frac{2\pi}{k}\mathcal{L}$, $\mathcal{W}\rightarrow -\frac{\pi}{2k} \mathcal{W}$; these normalization factors  ensure that the CFT on the boundary of this AdS theory  has naturally normalized OPE's. Here, however, these coefficients  simply complicate the expressions. Note that with this normalization, the connection $A$ takes the form
\begin{equation} \label{eq:spin2connectionAnorm}
\begin{split}
  A&=\left(e^\tau L_1+e^{-\tau} \mathcal{L}L_{-1}\right)dx+ L_0d\tau.
\end{split}
\end{equation}

What we have started with, then, is a 16-parameter (8 complex parameters) general connection. The equations of motion $da+a\wedge a$ fix $(w_{-2},w_{-1},w_0,w_1,l)$ in terms of $(\mathcal{L},\mathcal{W},w_2)$; the 16-parameter connection is reduced to a 6-parameter one. To simplify the equations, we assume all five of those complex parameters are independent of $x,\ol{x}$. The resulting connection is
\begin{equation} \label{eq:conna-3}
  \begin{split}
    a'&=\left(L_1+\mathcal{L}L_{-1}+\mathcal{W}W_{-2}\right)dx \\
    &\hspace{1cm}+\left\{wW_2+2\mathcal{L}wW_0+\mathcal{L}^2wW_{-2}-8\mathcal{W}wL_{-1}\right\}d\ol{x},
  \end{split}
\end{equation}
where we have replaced $w_2$ by $w$ for notational convenience. Returning to the big-$A$ connection via \eqref{eq:gaugeTransform}, we have
\begin{equation} \label{eq:connA-3}
  \begin{split}
    A'&=\left(e^\tau L_1+\mathcal{L}e^{-\tau}L_{-1}+\mathcal{W}e^{-2\tau}W_{-2}\right)dx \\
    &\hspace{1cm}+\left\{we^{2\tau}W_2+2\mathcal{L}wW_0+\mathcal{L}^2we^{-2\tau}W_{-2}-8\mathcal{W}we^{-\tau}L_{-1}\right\}d\ol{x}+L_0d\tau.
  \end{split}
\end{equation}

We wish to find a class of connections of the form \eqref{eq:connA-3} that have the same holonomy as \eqref{eq:spin2connectionA}, that remove the singularity from the spacetime, i.e. that change the volume of the space so that it is positive-definite. By doing this, we will show that any singularities in \eqref{eq:dsMetric} are not true singularities in the framework of the higher-spin theory as they are, in this case, gauge artifacts. We can do this by requiring that the $g_{zz}$ and $g_{\phi\phi}$ arising from \eqref{eq:connA-3} differ from those arising from \eqref{eq:spin2connectionA} by positive quantities. A more rigorous statement about singularities based on curvature invariants can be similarly developed  (see appendix (\ref{Physics}) for details of relevant calculations.).

  First let us write $A'$ in a simpler form:
  \begin{equation}
    A'=A+Xdx+Yd\ol{x}, \hspace{1cm} \ol{A'}=\ol{A}+\ol{X}d\ol{x}+\ol{Y}dx.
  \end{equation}
  The new veilbeins are then
  \begin{equation}
    e'=\frac{1}{2i}(A'-\ol{A'})=e+\frac{1}{2i}(Xdx-\ol{X}d\ol{x}+Yd\ol{x}-\ol{Y}dx),
  \end{equation}
  and the new metric components are
  \begin{equation}
  \begin{split}
    g'_{xx}&=g_{xx}+\frac{1}{5i}\Tr(e_x(X-\ol{Y}))-\frac{1}{20}\Tr((X-\ol{Y})^2), \\
    g'_{\ol{x}\ol{x}}&=g_{\ol{x}\ol{x}}-\frac{1}{20}\Tr((Y-\ol{X})^2), \\
    g'_{\ol{x}x}&=g_{\ol{x}x}+\frac{1}{10i}\Tr(e_x(Y-\ol{X}))-\frac{1}{20}\Tr((X-\ol{Y})(Y-\ol{X})).
  \end{split}
  \end{equation}
  To ensure $g'_{zz}-g_{zz}>0$ and $g'_{\phi\phi}-g_{\phi\phi}>0$, we can impose the following condition:
  \begin{equation}
  \begin{split}
    &\frac{1}{10i}\Tr(e_x(Y-\ol{X}))-\frac{1}{20}\Tr((X-\ol{Y})(Y-\ol{X}))- \\
    &\hspace{1cm}\left|\frac{1}{5i}\Tr(e_x(X-\ol{Y}))-\frac{1}{20}\Tr((X-\ol{Y})^2)+\Tr((Y-\ol{X})^2)\right|>0.
  \end{split}
  \end{equation}
  We also require that $A$ and $A'$ have the same holonomy. We impose the following conditions to this end:
  \begin{equation}
    \Tr(H[A']^2)=\Tr(H[A]^2), \hspace{1cm} \det(H[A'])=\det(H[A]).
  \end{equation}
  These are two complex conditions, which represent 4 degrees of freedom. We are left with a $6-4=2$-parameter family of connections that remove the cosmological singularity.

  What we started with was a 16-parameter general connection, and we end with a gauge-equivalent 2-parameter connection which gives rise to a singularity-free metric. The equations of motion fixed 10 parameters, and the two complex holonomy conditions fixed 4: $16-10-4=2$. The only assumption we have made is that all of the initial parameters are independent of $x$ and $\ol{x}$, and that was just for simplicity of the expressions; none of the real machinery depended on that assumption.

\section{Spin-4 Cosmology}\label{Spin4}
\subsection{Restricted connection}
  We now consider a simple extension of \eqref{eq:spin2connectionA}, along the same lines as~\cite{Krishnan:2013cra}, following a similar procedure to resolve the cosmological singularity. The $\sla(4,\mathbb{C})$-valued connection (see appendix \ref{sl4genapx} for explicit generators) we consider is
  \begin{equation} \label{eq:restConn}
    A'=A+Xdx,\hspace{1cm}X=\sum_{a=-2}^2 e^{a\tau}w_a(x,\ol{x},\tau)W_a+\sum_{a=-3}^3e^{a\tau}z_a(x,\ol{x},\tau)Z_a.
  \end{equation}
  To apply the equations of motion, we gauge transform this to a simple connection $a'$ via \eqref{eq:gaugeTransform}:
  \begin{equation}
    a'=\left(L_1+\frac{2\pi}{k}\mathcal{L}L_{-1}\right)dx+\left(\sum_{a=-2}^2 w_a(x,\ol{x},\tau)W_a+\sum_{a=-3}^3z_a(x,\ol{x},\tau)Z_a\right)dx.
  \end{equation}
  We require this connection to be flat, i.e. $da'+a'\wedge a'=0$, or in components, $\partial_\mu a'_\nu-\partial_\nu a'_\mu+[a'_\mu,a'_\nu]=0$. Since $a'$ has only one coordinate component, the commutator always vanishes, and so the equations of motion reduce to $\partial_\mu a'_\nu-\partial_\nu a'_\mu=0$. Only $a'_x$ is nonzero, so the equations of motion are
  \begin{equation}
    \partial_\tau a'_x=0, \hspace{1cm} \partial_{\ol{x}}a'_x=0.
  \end{equation}
  This means that $w_a$, $z_a$ are all independent of $\tau,\ol{x}$, but they are allowed to be arbitrary functions of $x$. So our flat connection is
  \begin{equation}
    a=\left(L_1+\frac{2\pi}{k}\mathcal{L}(x)L_{-1}\right)dx+Ydx,\hspace{1cm}Y=\sum_{a=-2}^2 w_a(x)W_a+\sum_{a=-3}^3z_a(x)Z_a.
  \end{equation}
  From now on we will suppress the $x$-dependence of $\mathcal{L}$, $w_a$, $z_a$.

  We can gauge-transform back to $A'$, which reads
  \begin{equation}
    A'=A+X dx,
  \end{equation}
  where
  \begin{equation}
    X=e^{-\tau L_0}Ye^{\tau L_0}=\sum_{a=-2}^2 e^{a\tau}w_aW_a+\sum_{a=-3}^3e^{a\tau}z_aZ_a.
  \end{equation}
  In order for the Einstein-Hilbert action in this theory to be real, it must be the case that $\ol{A}_aT^a=A^*_aT^a$. Thus our conjugate connection is
  \begin{equation}
    \ol{A}=\ol{A}_0+\ol{X}d\ol{x}, \hspace{1cm} \ol{X}=\sum_{a=-2}^2 e^{a\tau}\ol{w}_aW_a+\sum_{a=-3}^3e^{a\tau}\ol{z}_aZ_a
  \end{equation}
  Note that this differs from the conjugate connection in the $\sla(4,\mathbb{R})\times \sla(4,\mathbb{R})$ theory; in that theory, the generators $T^a$ in the conjugate connection are proportional to $e^{-a\rho}$ as opposed to $e^{a\tau}$.

  The new vielbein resulting from this connection is
  \begin{equation}
    e'=e+\frac{1}{2i}(Xdx-\ol{X}d\ol{x}),
  \end{equation}
  and so the new metric components are
  \begin{equation}
    g'_{xx}=g_{xx}-\frac{1}{20}\Tr(X^2), \hspace{1cm} g'_{\ol{x}\ol{x}}=g_{\ol{x}\ol{x}}-\frac{1}{20}\Tr(\ol{X}^2), \hspace{1cm} g'_{x\ol{x}}=g_{x\ol{x}}+\frac{1}{20}\Tr(X\ol{X}).
  \end{equation}
  Noting that $g_{\phi\phi}=g_{xx}+g_{\ol{x}\ol{x}}+g_{x\ol{x}}$ and $g_{zz}=-g_{xx}-g_{\ol{x}\ol{x}}+g_{x\ol{x}}$, we can see that sufficient conditions for $g'_{\phi\phi}-g_{\phi\phi}>0$ and $g'_{zz}-g_{zz}>0$ are
  \begin{equation} \label{eq:traceCond}
    \Tr(X^2)=\Tr(\ol{X}^2)=0, \hspace{1cm} \Tr(X\ol{X})>0.
  \end{equation}
  The first two conditions above in \ref{eq:traceCond} become
  \begin{equation} \label{eq:firstTwo}
  \begin{split}
    w_0^2-3w_{-1}w_1+12w_{-2}w_2+45z_0^2-120z_{-1}z_1+300z_{-2}z_2-1800z_{-3}z_3&=0, \\
    \ol{w}_0^2-3\ol{w}_{-1}\ol{w}_1+12\ol{w}_{-2}\ol{w}_2+45\ol{z}_0^2-120\ol{z}_{-1}\ol{z}_1+300\ol{z}_{-2}\ol{z}_2-1800\ol{z}_{-3}\ol{z}_3&=0.
  \end{split}
  \end{equation}
  These are equivalent, since the second is the conjugate of the first. The third condition in \ref{eq:traceCond} is
  \begin{equation}
    12 w_2\ol{w}_{-2}-3w_1\ol{w}_{-1}+w_0\ol{w}_0-1800z_3\ol{z}_{-3}+300 z_2\ol{z}_{-2}-120 z_1\ol{z}_{-1}+45z_0\ol{z}_0+\text{c.c.}>0.
  \end{equation}

  In addition to these conditions, we also require that $A'$ and $A$ have the same holonomy, so that they are related by a single-valued gauge transformation. To enforce this, we will equate the coefficients in the characteristic polynomials of the holonomy matrices of $A'$ and $A$. The holonomy matrix of a connection $A$ is given by
  \begin{equation}
    H[A]=\int A|_{z=const,\tau=0}=\int_0^{2\pi}d\phi\,A_\phi|_{\tau=0}.
  \end{equation}
  In order to simplify matters, we now take $w_a$, $z_a$, and $\mathcal{L}$ to be independent of $x$, so that they are all constants. Thus all of our connections are independent of $\phi$, and the holonomy matrix is
  \begin{equation}
    H[A]=2\pi A_\phi.
  \end{equation}

  For a general $4\times 4$ matrix $M$, the characteristic polynomial is
  \begin{equation}
  \begin{split}
    p_M(\lambda)&=\lambda^4-(\Tr M)\lambda^3+\frac{1}{2}((\Tr M)^2-\Tr(M^2))\lambda^2 \\
    &\hspace{1cm}-\frac{1}{6}((\Tr M)^3-3\Tr(M^2)(\Tr M)+2\Tr(M^3))\lambda+\det(M).
  \end{split}
  \end{equation}
  The matrices we consider belong to $\sla(4,\mathbb{C})$, and so are traceless, and this reduces to
  \begin{equation}
    p_M(\lambda)=\lambda^4-\frac{1}{2}\Tr(M^2)\lambda^2-\frac{1}{3}\Tr(M^3)\lambda+\det(M).
  \end{equation}
  We require then that
  \begin{equation}
    \Tr(H[A']^2)=\Tr(H[A]^2), \hspace{1cm} \Tr(H[A']^3)=\Tr(H[A]^3), \hspace{1cm} \det(H[A'])=\det(H[A]).
  \end{equation}
  The first condition is equivalent to \eqref{eq:firstTwo}. The second condition reads
  \begin{equation}
  \begin{split}
    0&=12 \mathcal{L}^2\pi^2 w_2+2k \mathcal{L}\pi(w_0+12 w_2z_{-1}-9w_1z_0+12w_0z_1-30 w_{-1}z_2+180w_{-2}z_3) \\
    &\hspace{1cm}-3k^2(10w_1z_{-2}-4w_0z_{-1}+3w_{-1}z_0-24(w_1z_{-1}z_0+w_{-1}z_1z_0+2w_0z_{-1}z_1) \\
    &\hspace{1cm}+24w_0z_0^2+120(w_1z_{-2}z_1+w_{-1}z_2z_{-1})+96(w_2z_{-1}^2+w_{-2}z_1^2)-240(w_2z_{-2}z_0+w_{-2}z_2z_0) \\
    &\hspace{1cm}+480(w_2z_{-3}z_1+w_{-2}z_3z_{-1})-60w_2z_{-3}-600(w_1z_{-3}z_2+w_{-1}z_3z_{-2}-2w_0z_{-3}z_3) \\
    &\hspace{1cm}-w_{-2}-4w_{-2}z_1).
  \end{split}
  \end{equation}
  The third condition is
  \begin{equation}
  \begin{split}
    0&=-1440 \mathcal{L}^3 \pi^3 z_3 - 12 k \mathcal{L}^2 \pi^2 (3 w_1^2-4 (2 w_0 w_2 - 3 z_1 + 12 z_1^2 - 30 z_0 z_2 + 60 z_{-1} z_3)) \\
    &\hspace{1cm}+4k^2 \mathcal{L}\pi(w_0^2 (48 z_1-1)+9(160 w_2^2 z_{-3} - 2 z_{-1} + 8 w_1^2 z_{-1} - 11 z_0^2 \\
    &\hspace{2cm}+8w_2(-5 w_1 z_{-2} + w_{-1} z_0)+32 z_{-1} z_1 + 48 z_0^2 z_1 - 128 z_{-1} z_1^2 - 100 z_{-2} z_2 \\
    &\hspace{2cm}-80 z_{-1} z_0 z_2 + 800 z_{-2} z_1 z_2 - 4000 z_{-3} z_2^2+4w_{-2}(w_2 - 8 w_2 z_1 + 10 w_1 z_2) \\
    &\hspace{2cm}+40 w_{-1}^2 z_3 + 800 z_{-3} z_3 + 1280 z_{-1}^2 z_3-3600 z_{-2} z_0 z_3 + 9600 z_{-3} z_1 z_3) \\
    &\hspace{1.5cm}+24 w_0 (2 w_2 z_{-1} - 3 w_1 z_0 - 5 (w_{-1} z_2 + 6 w_{-2} z_3))) \\
    &\hspace{1cm}+k^3(9 w_{-1}^2 (-1 + 4 w_1^2 + 16 z_1 - 64 z_1^2 + 960 z_{-1} z_3) \\
    &\hspace{1.5cm}+4(w_0^4+ 12 w_0 (6 w_2 (-8 z_{-1}^2 + 10 z_{-2} z_0 + 5 z_{-3} (-1 + 8 z_1)) \\
    &\hspace{3cm}-w_1 (-24 z_{-1} z_0 + z_{-2} (5 + 60 z_1) + 300 z_{-3} z_2)) \\
    &\hspace{2cm}+6 w_0^2 (z_{-1} (4 + 8 z_1) + 5 (-3 z_0^2 + 20 z_{-2} z_2 + 120 z_{-3} z_3)) \\
    &\hspace{2cm}+6 w_{-2} (4 w_0^2 w_2 + 6 (2 w_2 (-9 z_0^2 + 2 z_{-1} (-1 + 8 z_1))+w_1 (z_0 + 12 z_0 z_1 - 40 z_{-1} z_2)) \\
    &\hspace{3cm}+w_0 (1 + 4 z_1 - 96 z_1^2 + 120 z_0 z_2 + 480 z_{-1} z_3)) \\
    &\hspace{2cm}+ 144 w_{-2}^2 (w_2^2 + 5 (-5 z_2^2 + z_3 + 12 z_1 z_3)) \\
    &\hspace{2cm}+9(-400 w_2^2 z_{-2}^2 + 160 w_1 w_2 z_{-2} z_{-1} + 4 z_{-1}^2 -16 w_1^2 z_{-1}^2 - 10 z_{-2} z_0 + 24 z_{-1} z_0^2 \\
    &\hspace{3cm} + 81 z_0^4 - 64 z_{-1}^2 z_1 - 40 z_{-2} z_0 z_1 - 432 z_{-1} z_0^2 z_1 + 256 z_{-1}^2 z_1^2+960 z{-2} z_0 z_1^2 \\
    &\hspace{3cm} + 400 z_{-2} z_{-1} z_2 + 960 z_{-1}^2 z_0 z_2 - 1800 z_{-2} z_0^2 z_2 - 3200 z_{-2} z_{-1} z_1 z_2 \\
    &\hspace{3cm} + 10000 z_{-2}^2 z_2^2 - 2000 z_{-2}^2 z_3 - 3840 z_{-1}^3 z_3 + 14400 z_{-2} z_{-1} z_0 z_3 - 24000 z_{-2}^2 z_1 z_3 \\
    &\hspace{3cm}+5z_{-3}(-1 + 192 w_2^2 z_{-1} - 144 w_1 w_2 z_0 + 4 z_1 + 128 z_1^2 - 768 z_1^3 \\
    &\hspace{4cm} + w_1^2 (4 + 48 z_1) - 360 z_0 z_2 + 2880 z_0 z_1 z_2 - 4800 z_{-1} z_2^2 + 960 z_{-1} z_3 \\
    &\hspace{4cm}- 6480 z_0^2 z_3 + 11520 z_{-1} z_1 z_3))) \\
    &\hspace{1.5cm}-24w_{-1}(w_0^2 w_1 + 6 w_0 (z_0 - 8 z_0 z_1 + 20 z_{-1} z_2 + 100 z_{-2} z_3) \\
    &\hspace{2cm}+3(2 w_2 (-12 z_{-1} z_0 + 5 z_{-2} (-1 + 8 z_1))+w_1 (9 z_0^2 - 100 z_{-2} z_2)) \\
    &\hspace{3cm}+2 w_{-2} (2 w_1 w_2 + 5 (z_2 - 8 z_1 z_2 + 36 z_0 z_3))))).
  \end{split}
  \end{equation}

Summarizing, we started with a connection with 12 unspecified \emph{complex} constants ($w_a$, $z_a$), which represent 24 degrees of freedom. We have three complex conditions on these constants, which represent 6 degrees of freedom. We are thus left with an 18-parameter family of gauge transformations that eliminate the cosmological singularity.
Due to the generality of the above conditions we expect them to be of general type and therefore have the number of zeroes claimed above; we have checked that this is true in some limited cases.

\subsection{General connection and general singularity resolution}

We begin with an ansatz for an $\sla(4,\mathbb{C})$-valued connection which is a generalization of \eqref{eq:spin2connectionA}, in parallel with the literature for spin-4 black holes in $\AdS_3$ (note that this is the most general asymptotically $\dS_3$ connection):
\begin{equation}
\begin{split}
  a'&=\left(L_1+\mathcal{L}L_{-1}+\mathcal{W}W_{-2}+\mathcal{Z}Z_{-3}\right)dx \\
  &\hspace{1cm}+(w_{-2}W_{-2}+w_{-1}W_{-1}+w_0W_0+w_1W_1+w_2W_2+l L_{-1} \\
  &\hspace{2cm}+z_{-3}Z_{-3}+z_{-2}Z_{-2}+z_{-1}Z_{-1}+z_0Z_0+z_1Z_1+z_2Z_2+z_3Z_3)d\ol{x}
\end{split}
\end{equation}
where all of the coefficients of the generators belong to $\mathbb{C}$. Here we omit the normalization coefficients of $\mathcal{L}$, $\mathcal{W}$, and $\mathcal{Z}$ for simplicity. They can be restored by $\mathcal{L}\rightarrow \frac{2\pi}{k}\mathcal{L}$, $\mathcal{W}\rightarrow -\frac{\pi}{2k} \mathcal{W}$, and $\mathcal{Z}\rightarrow\frac{\pi}{720k}\mathcal{Z}$. With this normalization, the connection $A$ takes the form
\begin{equation} \label{eq:spin2connectionAnorm}
\begin{split}
  A&=\left(e^\tau L_1+e^{-\tau} \mathcal{L}L_{-1}\right)dx+ L_0d\tau.
\end{split}
\end{equation}

What we have started with, then, is a 32-parameter (16 complex parameters) general connection. The equations of motion $da+a\wedge a$ fix $(w_{-2},w_{-1},w_0,w_1,z_{-3},z_{-2},z_{-1},z_0,z_1,z_2,l)$ in terms of $(\mathcal{L},\mathcal{W},\mathcal{Z},w_2,z_3)$; the 32-parameter connection is reduced to a 10-parameter one. To simplify the equations, we assume all five of those complex parameters are independent of $x,\ol{x}$. The resulting connection is
\begin{equation} \label{eq:conna}
  \begin{split}
    a'&=\left(L_1+\mathcal{L}L_{-1}+\mathcal{W}W_{-2}+\mathcal{Z}Z_{-3}\right)dx \\
    &\hspace{1cm}+\left\{wW_2+\left(2\mathcal{L}w-200 \mathcal{W}z\right)W_0+\left(\left(\mathcal{L}^2-100\mathcal{Z}\right)w-100\mathcal{L}\mathcal{W}z\right)W_{-2}\right. \\
    &\hspace{1cm}+\left.zZ_3+3\mathcal{L}zZ_1+\left(3\mathcal{L}^2-60\mathcal{Z}\right)zZ_{-1}+\left(\mathcal{L}^3-44\mathcal{L}\mathcal{Z}\right)zZ_{-3}\right. \\
    &\hspace{1cm}\left.+\left(4320\mathcal{Z}z-8\mathcal{W}w\right)L_{-1}\right\}d\ol{x},
  \end{split}
\end{equation}
where we have replaced $w_2$ by $w$ and $z_3$ by $z$ for convenience. Returning to the big-$A$ connection via \eqref{eq:gaugeTransform}, we have
\begin{equation} \label{eq:connA}
  \begin{split}
    A'&=\left(e^\tau L_1+\mathcal{L}e^{-\tau}L_{-1}+\mathcal{W}e^{-2\tau}W_{-2}+\mathcal{Z}e^{-3\tau}Z_{-3}\right)dx \\
    &\hspace{1cm}+\left\{we^{2\tau}W_2+\left(2\mathcal{L}w-200 \mathcal{W}z\right)W_0+\left(\left(\mathcal{L}^2-100\mathcal{Z}\right)w-100\mathcal{L}\mathcal{W}z\right)e^{-2\tau}W_{-2}\right. \\
    &\hspace{1cm}+\left.ze^{3\tau}Z_3+3\mathcal{L}ze^\tau Z_1+\left(3\mathcal{L}^2-60\mathcal{Z}\right)ze^{-\tau}Z_{-1}+\left(\mathcal{L}^3-44\mathcal{L}\mathcal{Z}\right)ze^{-3\tau}Z_{-3}\right. \\
    &\hspace{1cm}\left.+\left(4320\mathcal{Z}z-8\mathcal{W}w\right)e^{-\tau}L_{-1}\right\}d\ol{x}+L_0d\tau.
  \end{split}
\end{equation}

We wish to find a class of connections of the form \eqref{eq:connA} that have the same holonomy as \eqref{eq:spin2connectionA}, and take the metric \eqref{eq:dsMetric} with singularities, to a metric without singularities. By this, we mean that we hope to make all components of the metric non-vanishing, to show that any singularities in \eqref{eq:dsMetric} are not true singularities as they can be eliminated via a holonomy-preserving gauge transformation. We can do this by requiring that the $g_{zz}$ and $g_{\phi\phi}$ arising from \eqref{eq:connA} differ from those arising from \eqref{eq:spin2connectionA} by positive quantities.

  First let us write $A'$ in a simpler form:
  \begin{equation}
    A'=A+Xdx+Yd\ol{x} \hspace{1cm} \ol{A'}=\ol{A}+\ol{X}d\ol{x}+\ol{Y}dx.
  \end{equation}
  The new veilbeins are then
  \begin{equation}
    e'=\frac{1}{2i}(A'-\ol{A'})=e+\frac{1}{2i}(Xdx-\ol{X}d\ol{x}+Yd\ol{x}-\ol{Y}dx),
  \end{equation}
  and the new metric components are
  \begin{equation}
  \begin{split}
    g'_{xx}&=g_{xx}+\frac{1}{5i}\Tr(e_x(X-\ol{Y}))-\frac{1}{20}\Tr((X-\ol{Y})^2), \\
    g'_{\ol{x}\ol{x}}&=g_{\ol{x}\ol{x}}-\frac{1}{20}\Tr((Y-\ol{X})^2), \\
    g'_{\ol{x}x}&=g_{\ol{x}x}+\frac{1}{10i}\Tr(e_x(Y-\ol{X}))-\frac{1}{20}\Tr((X-\ol{Y})(Y-\ol{X})).
  \end{split}
  \end{equation}
  To ensure $g'_{zz}-g_{zz}>0$ and $g'_{\phi\phi}-g_{\phi\phi}>0$, we can impose the following condition:
  \begin{equation}
  \begin{split}
    &\frac{1}{10i}\Tr(e_x(Y-\ol{X}))-\frac{1}{20}\Tr((X-\ol{Y})(Y-\ol{X}))- \\
    &\hspace{1cm}\left|\frac{1}{5i}\Tr(e_x(X-\ol{Y}))-\frac{1}{20}\Tr((X-\ol{Y})^2)+\Tr((Y-\ol{X})^2)\right|>0.
  \end{split}
  \end{equation}
  We also require that $A$ and $A'$ have the same holonomy. As before, we can impose the following conditions to this end:
  \begin{equation}
  \label{Eq:CondHol-4}
    \Tr(H[A']^2)=\Tr(H[A]^2), \hspace{1cm} \Tr(H[A']^3)=\Tr(H[A]^3), \hspace{1cm} \det(H[A'])=\det(H[A]).
  \end{equation}
  The first condition in equation \ref{Eq:CondHol-4} reads explicitly as
  \begin{equation}
    0=4 \mathcal{L}^2 w^2-720 \mathcal{L}^3 z^2+10000 \mathcal{W}^2 z^2+\mathcal{W}w\left(\frac{11}{2} -500 \mathcal{L}z\right)-\mathcal{Z}z\left(1800-25200 \mathcal{L}z\right)-300 \mathcal{Z}w^2
  \end{equation}
  The second condition in equation \ref{Eq:CondHol-4} reads
  \begin{equation}
  \begin{split}
    0&=18432 \mathcal{L}^4 w z^2 + 3 \mathcal{W} (-1 + 19200 (w^2 - 15 z) z \mathcal{Z}) + 120 w \mathcal{Z} (1 + 288000 z^2 \mathcal{Z}) \\
    &\hspace{1cm} + 8 \mathcal{L}^2 (36 \mathcal{W}z (8 w^2 + 149 z) - w (1 + 322560 z^2 \mathcal{Z})) \\
    &\hspace{1cm} + 4 \mathcal{L} (-50400 w \mathcal{W}^2 z^2 + 2160 w z \mathcal{Z} + \mathcal{W} (16 w^2 + 71 z + 36000000 z^3 \mathcal{Z})) \\
    &\hspace{1cm} - 32 w \mathcal{W}^2 (6 w^2 + 5 z) - 576 \mathcal{L}^3 z (w + 2000 \mathcal{W} z^2).
  \end{split}
  \end{equation}
  The third condition in equation \ref{Eq:CondHol-4} reads
  \begin{equation}
  \begin{split}
    0&=2985984 \mathcal{L}^6 z^4 + 1600000000 \mathcal{W}^4 z^4 + 18432 \mathcal{L}^5 z^2 (5 w^2 + 18 z) \\
    &\hspace{1cm} + 32 \mathcal{L}^4 (8 w^4 + 207 z^2 - 401472 w \mathcal{W} z^3 + 15863040 z^4 \mathcal{Z}) \\
    &\hspace{1cm} + 12 \mathcal{W}^2 (3 w^2 - 40 z + 32000 z^2 (11 w^2 + 900 z) \mathcal{Z}) \\
    &\hspace{1cm} + 45 \mathcal{Z} (-1 + 6400 (5 w^4 + 4 w^2 z + 90 z^2) \mathcal{Z} - 82944000000 z^4 \mathcal{Z}^2) \\
    &\hspace{1cm} + \mathcal{L} (-125440000 w \mathcal{W}^3 z^3 + 3 w \mathcal{W} (7 + 1280 (866 w^2 - 3615 z) z \mathcal{Z}) \\
    &\hspace{2cm} + 120 \mathcal{Z} (-25 w^2 + 96 z + 57600 (131 w^2 - 480 z) z^2 \mathcal{Z}) \\
    &\hspace{2cm}+ 32 \mathcal{W}^2 (24 w^4 - 586 w^2 z + 1235 z^2 - 900000000 z^4 \mathcal{Z})) \\
    &\hspace{1cm} - 8 (160 w \mathcal{W}^3 (42 w^2 - 425 z) z +
    \mathcal{L}^3 (-5 w^2 + 18 z + 32 w \mathcal{W} (250 w^2 - 63 z) z \\
    &\hspace{3cm} - 61200000 \mathcal{W}^2 z^4 + 23040 (55 w^2 - 81 z) z^2 \mathcal{Z}) \\
    &\hspace{2cm} + 120 w \mathcal{W} \mathcal{Z} (21 w^2 - 65 z + 36000000 z^3 \mathcal{Z}) \\
    &\hspace{2cm} + \mathcal{L}^2 (-96 \mathcal{W}^2 z^2 (5747 w^2 + 1725 z) + w \mathcal{W} (-8 w^2 + 409 z - 167904000 z^3 \mathcal{Z}) \\
    &\hspace{3cm} + 60 \mathcal{Z} (80 w^4 - 480 w^2 z + 1779 z^2 + 27216000 z^4 \mathcal{Z}))).
  \end{split}
  \end{equation}
The actual form of the conditions is not very  illuminating, but we have included them here for completeness. What is important is that the conditions are of general type which allows us to proceed with a general counting as follows. What we started with was a 32-parameter general connection, and we end with a gauge-equivalent 4-parameter connection which gives rise to a singularity-free metric. The equations of motion fixed 22 parameters, and the three complex holonomy conditions fixed 6: $32-22-6=4$. The only assumption we have made is that all of the initial parameters are independent of $x$ and $\ol{x}$, and that was just for simplicity of the expressions; none of the real machinery depended on that assumption.

\section{Resolution of cosmological singularity in spin-N cosmology}\label{SpinN}
  Now to return to the restricted connection: it seems that this specific choice of restricted connection can be generalized to spin-$N$ dS gravity. Such a generalized connection might take the form
  \begin{equation}
    A'=A+Xdx,\hspace{1cm} X=\sum_{s=3}^{N}\sum_{a=-(s-1)}^{s-1} e^{a\tau}w^{(s)}_a W^{(s)}_a,
  \end{equation}
  where $W^{(s)}_a$ is an $\sla(N)$ generator of weight $a$ in the spin-$s$ multiplet of the principal embedding of $\sla(2)$ into $\sla(N)$, and $w^{(s)}_a$ are simply the corresponding complex coefficients. It is easy to check that this reduces to \eqref{eq:restConn} for $N=4$, with $W^{(3)}_a=W_a$ and $W^{(4)}_a=Z_a$.

  For the spin-4 case, we had $5+7=12$ complex constants; for spin-$N$, there will be $5+7+\cdots+(2N-1)=N^2-4$ complex constants representing $2N^2-8$ degrees of freedom. The conditions \eqref{eq:traceCond} will be unchanged, and will always represent 2 degrees of freedom. The holonomy matrices in spin-$N$ gravity will have $N$ eigenvalues to fix (this is equivalent to fixing the coefficients of the characteristic polynomial); one will be fixed by the intrinsic tracelessness of the matrix, and one will be fixed by the conditions \eqref{eq:traceCond}, leaving $N-2$ complex conditions, or $2N-4$ degrees of freedom. Thus for spin-$N$ gravity, there will be a $(2N^2-8)-(2N-4)-2=2(N^2-N-3)$-parameter family of gauge transformations that remove the cosmological singularity. We can not explicitly check that the conditions are of general type. However, based on the structure of the spin-3 and spin-4 case, we  assume that the restrictions are of general type and that our counting of degrees of freedom in the equations is  accurate.

\section{Conclusions}\label{Conclusions}
In this work we have explored the possibility of resolving cosmological singularities in the context of higher-spin 3d gravity. We have explicitly demonstrated that a mechanism proposed in the case of spin-3 can be generalized to spin-4. We have also included a resolution which is general in that it requires only that the asymptotic form of dS${}_3$ be preserved. We have also provided evidence that the mechanism works for the general spin-N case.

In the appendix we have discussed some of the aspects of the geometry from the diffeomorphic (spin-2) invariant point of view. For example, we have shown that the minimal volume of the cosmological evolution depends on all the parameters of the gauge transformation. Namely, the minimal volume depends on all the charges involved in the construction of higher-spin cosmology. From this point of view, and from our discussion of the asymptotic structure in the presence of higher spin fields in section \ref{hsc-hsbh}, it seems that cosmology is mildly modified by the presence of higher spin fields as compared to the situation in higher-spin black holes.

Our work is a step in the study of cosmological singularities in higher-spin gravity theories. It would be interesting to develop more general criteria for characterizing generic singularities in higher-spin theories.

Given that some cosmological horizons in asymptotically dS${}_3$ can be attributed an entropy, an interesting open problem is the consideration of thermodynamics in this context  along the lines of  \cite{Park:1998qk,Balasubramanian:2001nb}. Indeed, as this paper was being prepared for posting a preprint appeared \cite{Krishnan:2013zya} that discusses some such aspects in the presence of higher spin fields. We hope to return to some of these problems in the future.

\section*{Acknowledgments}
L. PZ thanks Alejandro Cabo-Bizet and Soo-Jong Rey and acknowledges the hospitality at APCTP, Korea and ICTP, Italy  where part of this work was done. This research was supported in part by the Department of Energy under grant DE-FG02-95ER40899 to the University of Michigan. BB thanks the MCTP for hospitality during the initial stages of this project. The work of BB has been financially supported under a research grant from Troy University. NR is thankful to the Honors Summer Fellowship program for undergraduate students at the University of Michigan for generous support.

\appendix

\section{Algebra: Generators and some properties}\label{Generators}
In this appendix we present the explicit realizations of the algebra generators that we used in the main body of the paper.

\subsection{$\sla(2)$ generators}
\label{sl2genapx}
We use the following matrices to furnish a basis of $\sla(2)$:

\begin{eqnarray}
  L_1&=&\begin{pmatrix}
    0 & 0 \\
    1 & 0 \end{pmatrix}, \qquad
  L_0= \begin{pmatrix}
    1/2 & 0 \\
    0 & -1/2 \end{pmatrix}, \qquad
  L_{-1}=\begin{pmatrix}
    0 & -1 \\
    0 & 0 \end{pmatrix}
\end{eqnarray}

The commutation relations satisfied by these generators is
\begin{equation}
\begin{split}
  [L_m,L_n]&=(m-n)L_{m+n}
\end{split}
\end{equation}

\subsection{$\sla(3)$ generators}
\label{sl3genapx}
We use the following matrices to furnish a basis of $\sla(3)$:

\begin{eqnarray}
  L_1&=&\begin{pmatrix} 0 & 0 & 0  \\ 1 & 0 & 0 \\
0 & 1 & 0 \end{pmatrix}, \qquad
  L_0= \begin{pmatrix} 1 & 0 & 0 \\ 0 & 0 & 0  \\
0 & 0 & -1  \end{pmatrix}, \qquad
  L_{-1}=\begin{pmatrix}
0 & -2 & 0 \\
0 & 0 & -2  \\
0 & 0 & 0
  \end{pmatrix} \nn \\
  W_{1}&=&\begin{pmatrix}
0 & 0 & 0 \\
1 & 0 & 0  \\
0 & -1 & 0 \end{pmatrix}, \qquad
W_{0}= \begin{pmatrix}
2/3 & 0 & 0 \\
0 & -4/3 & 0  \\
0 & 0 & 2/3 \end{pmatrix}, \qquad
W_{1}= \begin{pmatrix}
0 & -2 & 0 \\
0 & 0 & 2 \\
0 & 0 & 0  \end{pmatrix} \\
W_2 &=& \begin{pmatrix}
0 & 0 & 0 \\
0 & 0 & 0 \\
2 & 0 & 0 \end{pmatrix}, \qquad
W_{-2}= \begin{pmatrix}
0 & 0 & 8 \\
0 & 0 & 0 \\
0 & 0 & 0 \end{pmatrix} \nn
\end{eqnarray}

The commutation relations satisfied by these generators is
\begin{equation}
\begin{split}
  [L_m,L_n]&=(m-n)L_{m+n} \\
  [L_m, W_p]&=(2m-p)W_{m+p} \\
  [W_p, W_q]&=-\frac{1}{3}(p-q)(2p^2 + 2q^2-pq-8)L_{p+q}
\end{split}
\end{equation}

\subsection{$\sla(4)$ generators}
\label{sl4genapx}
We use the following matrices to furnish a basis of $\sla(4)$:

\begin{eqnarray}
  L_1&=&\begin{pmatrix}
  0 & 0 & 0 & 0 \\
  1 & 0 & 0 & 0 \\
  0 & 1 & 0 & 0 \\
  0 & 0 & 1 & 0 \end{pmatrix}, \qquad
  L_0= \begin{pmatrix}
  \frac{3}{2} & 0 & 0 & 0 \\
  0 & \frac{1}{2} & 0 & 0 \\
  0 & 0 & -\frac{1}{2}& 0  \\
  0 & 0 & 0 & -\frac{3}{2} \end{pmatrix}, \qquad
  L_{-1}=\begin{pmatrix}
0 & -3 & 0 & 0 \\
0 & 0 & -4 & 0  \\
0 & 0 & 0 & -3 \\
0 & 0 & 0 & 0  \end{pmatrix} \nn \\
Z_{-1} &=& \begin{pmatrix}
0 & 24 & 0 & 0 \\
0 & 0 & -48 & 0 \\
0 & 0 & 0 & 24 \\
0 & 0 & 0 & 0 \end{pmatrix}, \qquad
Z_{0}=\begin{pmatrix}
-6 & 0 & 0 & 0 \\
0 & 18 & 0 & 0 \\
0 & 0 & -18 & 0 \\
0 & 0 & 0 & 6 \end{pmatrix}, \qquad
Z_{1}= \begin{pmatrix}
0 & 0 & 0 & 0 \\
-8 & 0 & 0 & 0 \\
0 & 12 & 0 & 0 \\
0 & 0 & -8 & 0 \end{pmatrix} \nn \\
Z_{-2}&=& \begin{pmatrix}
0 & 0 & -120 & 0 \\
0 & 0 & 0 & 120 \\
0 & 0 & 0 & 0 \\
0 & 0 & 0 & 0 \end{pmatrix}, \qquad
Z_{2}= \begin{pmatrix}
0 & 0 & 0 & 0 \\
0 & 0 & 0 & 0 \\
-10 & 0 & 0 & 0 \\
0 & 10 & 0 & 0 \end{pmatrix} \\
Z_{-3}&=&\begin{pmatrix}
0 & 0 & 0 & 720 \\
0 & 0 & 0 & 0 \\
0 & 0 & 0 & 0 \\
0 & 0 & 0 & 0 \end{pmatrix}, \qquad
Z_3 = \begin{pmatrix}
0 & 0 & 0 & 0 \\
0 & 0 & 0 & 0 \\
0 & 0 & 0 & 0 \\
-20 & 0 & 0 & 0 \end{pmatrix} \nn \\
W_1&=& \begin{pmatrix}
0 & 0 & 0 & 0 \\
2 & 0 & 0 & 0 \\
0 & 0 & 0 & 0 \\
0 & 0 & -2 & 0 \end{pmatrix}, \qquad
W_0 = \begin{pmatrix}
2 & 0 & 0 & 0 \\
0 & -2 & 0 & 0 \\
0 & 0 & -2 & 0 \\
0 & 0 & 0 & 2 \end{pmatrix}, \qquad
W_{-1}= \begin{pmatrix}
0 & -6 & 0 & 0 \\
0 & 0 & 0 & 0 \\
0 & 0 & 0 & 6 \\
0 & 0 & 0 & 0 \end{pmatrix} \nn \\
W_{-2}&=&  \begin{pmatrix}
0 & 0 & 24 & 0 \\
0 & 0 & 0 & 24 \\
0 & 0 & 0 & 0 \\
0 & 0 & 0 & 0 \end{pmatrix}, \qquad
W_2 = \begin{pmatrix}
0 & 0 & 0 & 0 \\
0 & 0 & 0 & 0 \\
2 & 0 & 0 & 0 \\
0 & 2 & 0 & 0 \end{pmatrix} \nn
\end{eqnarray}

In the above basis, we find that the commutation relations are
\begin{eqnarray}
  \, [L_m, L_n]&=&(m-n)L_{m+n} \nn \\
\, [L_m, W_n]&=&(2m-n) W_{m+n} \nn \\
\, [L_m, Z_n]&=&(3m-n) Z_{m+n}  \\
\, [W_m, W_n]&=& -\frac{2}{5} (m-n) Z_{m+n} \nn \\
&& +\frac{4}{25}(m-n)\bigg(n^4+m^4+n^3m+nm^3+n^2m^2 \nn \\
&& \kern12em -15(n^2+m^2)+44\bigg)L_{m+n} \nn \\
\, [W_m, Z_n]&=&(10m^3-10m^2n+6mn^2-2n^3-34m+18n)W_{m+n}\nn \\
\, [Z_m, Z_n]&=& -(m-n)(2m^2+2n^2-2mn-14)Z_{m+n} \nn \\
&& + 4(m-n)\bigg(3m^4+3n^4-2(m^3n+mn^3)+4n^2m^2\nn \\
&& \kern10em -39(m^2+n^2)+20mn+108\bigg) L_{m+n}\nn
\end{eqnarray}
and the traces of combinations of generators are
\begin{equation}
\begin{split}
  \Tr(L_{-1}L_1)&=-10 \\
  \Tr(L_0L_0)&=5 \\
  \Tr(W_{-2}W_2)&=96 \\
  \Tr(W_{-1}W_1)&=-24 \\
  \Tr(W_0W_0)&=16 \\
  \Tr(Z_{-3}Z_3)&=-14400 \\
  \Tr(Z_{-2}Z_2)&=2400 \\
  \Tr(Z_{-1}Z_1)&=-960 \\
  \Tr(Z_0Z_0)&=720
\end{split}
\end{equation}
All other traces vanish.

\subsection{Weights and the structure of traces}

In an algebra, we can have an operator $L_0$, under which all other operators have a weight
\begin{equation}
[L_0,A_{n}]= nA_n.
\end{equation}
The algebra therefore divides up into a direct sum of the different weight spaces under this operator.

In a representation of the algebra, the trace will only be non zero if the sum of the weights under this $L_0$ operator are zero.  This is always the case because
\begin{eqnarray}
&&[L_0, A_{n} B_{m}]= [L_0,A_n] B_m+A_n[L_0, B_m]=(n+m)A_n B_m \\
&&[L_0, A_{n} B_{m}]= (n+m)A_n B_m
\end{eqnarray}

Taking the trace of both sides, we get
\begin{equation}
0={\rm Tr}([L_0, A_{n} B_{m}])=(n+m){\rm Tr}(A_n B_n).
\end{equation}
Thus, the combination $(n+m){\rm Tr}(A_n B_n)=0$.  This means that if ${\rm Tr}(A_n B_n)\neq 0$, then $(n+m)=0$, i.e., for the product to have non-vanishing  trace, the sum of the weights under the $L_0$ operator must be 0.

This must be true for all such $L_0$ operators in the algebra, for example, all of the Cartan subalgebra generators.  Hence, for the trace to be non zero, the weight w.r.t. any given Cartan element must be zero,  i.e., ${\rm Tr}(W_{\vec{a}} W_{\vec{b}})\neq 0 \rightarrow \vec{a}+\vec{b}=0$, where the vectors $\vec{a}$ are vectors of weights under the Cartan generators $\vec{H}$.

\section{Some diffeomorphic-invariant information}\label{Physics}
In this section we examine some of the properties of the geometry and how the gauge transformation affects them. In particular we check that the minimal volume in the cosmological evolution is, in fact, determined by all the charges of the gauge transformation.

The dS${}_3$ analog of the BTZ black hole is
\begin{equation}
  ds^2=-dt^2+\left(e^t d\ol{x}+\frac{2\pi}{k}\mathcal{L}e^{-t}dx\right)\left(e^t dx+\frac{2\pi}{k}\ol{\mathcal{L}}e^{-t}d\ol{x}\right)
\end{equation}
We will call this metric $g$.

The metric arising from the (restricted) $\SL(4,\mathbb{C})$ connection is
\begin{equation}
\begin{split}
  ds^2&=-dt^2+\left(e^t d\ol{x}+\frac{2\pi}{k}\mathcal{L}e^{-t}dx\right)\left(e^t dx+\frac{2\pi}{k}\ol{\mathcal{L}}e^{-t}d\ol{x}\right)-\frac{4}{5} (w_0 dx -  \ol{w}_0d\ol{x})^2 \\
  &\hspace{1cm}+\frac{12}{5} (w_{-1} dx - \ol{w}_{-1} d\ol{x}) (w_1 dx -
   \ol{w}_1 d\ol{x}) - \frac{48}{5} (w_{-2}dx - \ol{w}_{-2}d\ol{x})(w_2 dx - \ol{w}_2 d\ol{x}) \\
   &\hspace{1cm} - 36 (z_0 dx - \ol{z}_0 d\ol{x})^2 + 96 (z_{-1}dx - \ol{z}_{-1}d\ol{x}) (z_1 dx - \ol{z}_1 d\ol{x}) \\
   &\hspace{1cm} - 240 (z_{-2}dx - \ol{z}_{-2}d\ol{x}) (z_2 dx - \ol{z}_2 d\ol{x}) + 1440 (z_{-3}dx - \ol{z}_{-3}d\ol{x}) (z_3 dx - \ol{z}_3 d\ol{x})
\end{split}
\end{equation}
We will call this metric $g'$.

In the process of resolving the cosmological singularity in $g$, we enforced that $g'_{xx}=g_{xx}$ and $g'_{\ol{x}\ol{x}}=g_{\ol{x}\ol{x}}$. Thus we can rewrite $g'$ as
\begin{equation}
  ds^2=-dt^2+\left(e^t d\ol{x}+\frac{2\pi}{k}\mathcal{L}e^{-t}dx\right)\left(e^t dx+\frac{2\pi}{k}\ol{\mathcal{L}}e^{-t}d\ol{x}\right)+Mdxd\ol{x},
\end{equation}
where
\begin{equation}
  M=\frac{1}{20}(12 w_2\ol{w}_{-2}-3w_1\ol{w}_{-1}+w_0\ol{w}_0-1800z_3\ol{z}_{-3}+300 z_2\ol{z}_{-2}-120 z_1\ol{z}_{-1}+45z_0\ol{z}_0+c.c.)
\end{equation}
We also enforce that $M>0$.

We would like to examine some physical properties of this metric. We start by calculating the determinant of the metrics:
\begin{equation}
  \det g=\left(\frac{1}{2}e^{2t}-\frac{2\pi^2}{k^2}\mathcal{L}\ol{\mathcal{L}}e^{-2t}\right)^2
\end{equation}
\begin{equation}
\begin{split}
  \det g'&=\left(\frac{1}{2}e^{2t}+\frac{2\pi^2}{k^2}\mathcal{L}\ol{\mathcal{L}}e^{-2t}+\frac{1}{2}M\right)^2-\frac{4\pi^2}{k^2}\mathcal{L}\ol{\mathcal{L}} \\
  &=\left(\frac{1}{2}e^{2t}-\frac{2\pi^2}{k^2}\mathcal{L}\ol{\mathcal{L}}e^{-2t}+\frac{1}{2}M\right)^2+\frac{4\pi^2}{k^2}\mathcal{L}\ol{\mathcal{L}}Me^{-2t}
\end{split}
\end{equation}
We are reassured that we have removed the singularity, since clearly $\det g'>\det g$, and both are non-negative quantities.

The volume of our space is proportional to the square root of $|\det g'|$. Since $\det g'$ is a positive definite quantity, the volume is simply proportional to $\sqrt{\det g'}$. We can calculate the minimum volume (by minimizing $\det g'$):
\begin{equation}
\begin{split}
  \frac{d}{dt}(\det g')=0&=\left(e^{2t}+\frac{4\pi^2}{k^2}\mathcal{L}\ol{\mathcal{L}}e^{-2t}+M\right)\left(e^{2t}-\frac{4\pi^2}{k^2}\mathcal{L}\ol{\mathcal{L}}e^{-2t}\right)
\end{split}
\end{equation}
The quantity in the left parentheses is the sum of three positive definite terms, and can never be zero. So the minimum occurs when
\begin{equation}
  e^{4t}=\frac{4\pi^2}{k^2}\mathcal{L}\ol{\mathcal{L}}
\end{equation}
We know it is a minimum because
\begin{equation}
  \frac{d^2}{dt^2}(\det g')=\frac{1}{2}\left(2e^{2t}-\frac{8\pi^2}{k^2}\mathcal{L}\ol{\mathcal{L}}e^{-2t}\right)^2+\left(e^{2t}+\frac{4\pi^2}{k^2}\mathcal{L}\ol{\mathcal{L}}e^{-2t}+M\right)\left(2e^{2t}+\frac{8\pi^2}{k^2}\mathcal{L}\ol{\mathcal{L}}e^{-2t}\right)
\end{equation}
is a sum of positive-definite quantities.

The minimum volume is then
\begin{equation}
  \sqrt{\det g'}|_{e^{4t}=\frac{4\pi^2}{k^2}\mathcal{L}\ol{\mathcal{L}}}=\sqrt{\frac{M^2}{4}+\frac{2\pi}{k}M\sqrt{\mathcal{L}\ol{\mathcal{L}}}}
\end{equation}
We can see that as $M$ (which consists of constants from the higher-spin gauge transformation only) goes to zero, the minimum volume goes to zero, and the singularity is restored. The minimum volume depends both on the initial charges deforming dS${}_3$, $\mathcal{L}$ and $\ol{\mathcal{L}}$, and all of the higher-spin parameters, through $M$.

The scalar curvature is given by
\begin{equation}
\begin{split}
  2 (Ric)&=(3k^4e^{8\tau}+10k^4 Me^{6\tau}+11k^4M^2 e^{4\tau}+4k^4 M^3 e^{2\tau}-48k^2\pi^2 \mathcal{L}\ol{\mathcal{L}}e^{4\tau}+88k^2\pi^2 M \mathcal{L}\ol{\mathcal{L}}e^{2\tau} \\
  &\hspace{2cm}+104k^2\pi^2 M^2 \mathcal{L}\ol{\mathcal{L}}+16k^2\pi^2 M^3 \mathcal{L}\ol{\mathcal{L}}e^{-2\tau}+288\pi^4 \mathcal{L}^2\ol{\mathcal{L}}^2+352\pi^4 M \mathcal{L}^2\ol{\mathcal{L}}^2e^{-2\tau} \\
  &\hspace{2cm}+176\pi^4 M^2 \mathcal{L}^2\ol{\mathcal{L}}^2e^{-4\tau}-768k^{-2}\pi^6 \mathcal{L}^3\ol{\mathcal{L}}^3e^{-4\tau}+640k^{-2}\pi^6M \mathcal{L}^3\ol{\mathcal{L}}^3e^{-6\tau} \\
  &\hspace{2cm} + 768k^{-4}\pi^8 \mathcal{L}^4\ol{\mathcal{L}}^4e^{-8\tau}) \\
  &\hspace{1cm}/\left[\left(\frac{1}{2}e^{2t}-\frac{2\pi^2}{k^2}\mathcal{L}\ol{\mathcal{L}}e^{-2t}+\frac{1}{2}M\right)^2+\frac{4\pi^2}{k^2}\mathcal{L}\ol{\mathcal{L}}Me^{-2t}\right]^2
\end{split}
\end{equation}
The denominator above is the square of the determinant of the metric, which we have shown to be non-vanishing everywhere. Thus the scalar curvature is finite everywhere. Similarly, the denominator of the Kretschmann scalar $R^{abcd}R_{abcd}$ is the fourth power of $\det g'$, and so is finite everywhere.


\bibliographystyle{JHEP}
\bibliography{RGHS}

\providecommand{\href}[2]{#2}\begingroup\raggedright\begin{thebibliography}{10}

\bibitem{Vasiliev:1999ba}
M.~A. Vasiliev, {\it {Higher spin gauge theories: Star product and AdS space}},
   \href{http://xxx.lanl.gov/abs/hep-th/9910096}{{\tt hep-th/9910096}}.

\bibitem{Bekaert:2005vh}
X.~Bekaert, S.~Cnockaert, C.~Iazeolla, and M.~Vasiliev, {\it {Nonlinear higher
  spin theories in various dimensions}},
  \href{http://xxx.lanl.gov/abs/hep-th/0503128}{{\tt hep-th/0503128}}.

\bibitem{Sezgin:2002rt}
E.~Sezgin and P.~Sundell, {\it {Massless higher spins and holography}},  {\em
  Nucl.Phys.} {\bf B644} (2002) 303--370,
  [\href{http://xxx.lanl.gov/abs/hep-th/0205131}{{\tt hep-th/0205131}}].

\bibitem{HaggiMani:2000ru}
P.~Haggi-Mani and B.~Sundborg, {\it {Free large N supersymmetric Yang-Mills
  theory as a string theory}},  {\em JHEP} {\bf 0004} (2000) 031,
  [\href{http://xxx.lanl.gov/abs/hep-th/0002189}{{\tt hep-th/0002189}}].

\bibitem{Beisert:2004di}
N.~Beisert, M.~Bianchi, J.~F. Morales, and H.~Samtleben, {\it {Higher spin
  symmetry and N=4 SYM}},  {\em JHEP} {\bf 0407} (2004) 058,
  [\href{http://xxx.lanl.gov/abs/hep-th/0405057}{{\tt hep-th/0405057}}].

\bibitem{Gross:1987ar}
D.~J. Gross and P.~F. Mende, {\it {String Theory Beyond the Planck Scale}},
  {\em Nucl.Phys.} {\bf B303} (1988) 407.

\bibitem{Gross:1988ue}
D.~J. Gross, {\it {High-Energy Symmetries of String Theory}},  {\em
  Phys.Rev.Lett.} {\bf 60} (1988) 1229.

\bibitem{Blencowe:1988gj}
M.~Blencowe, {\it {A Consisten Interacting Massless Higher Spin Field Theory in
  $D = (2+1)$}},  {\em Class.Quant.Grav.} {\bf 6} (1989) 443.

\bibitem{Bordemann:1989zi}
M.~Bordemann, J.~Hoppe, and P.~Schaller, {\it {Infinite Dimensional Matrix
  Algebras}},  {\em Phys.Lett.} {\bf B232} (1989) 199.

\bibitem{Henneaux:2010xg}
M.~Henneaux and S.-J. Rey, {\it {Nonlinear $W_{infinity}$ as Asymptotic
  Symmetry of Three-Dimensional Higher Spin Anti-de Sitter Gravity}},  {\em
  JHEP} {\bf 1012} (2010) 007, [\href{http://xxx.lanl.gov/abs/1008.4579}{{\tt
  arXiv:1008.4579}}].

\bibitem{Campoleoni:2010zq}
A.~Campoleoni, S.~Fredenhagen, S.~Pfenninger, and S.~Theisen, {\it {Asymptotic
  symmetries of three-dimensional gravity coupled to higher-spin fields}},
  {\em JHEP} {\bf 1011} (2010) 007,
  [\href{http://xxx.lanl.gov/abs/1008.4744}{{\tt arXiv:1008.4744}}].

\bibitem{Vasiliev:1990en}
M.~A. Vasiliev, {\it {Consistent equation for interacting gauge fields of all
  spins in (3+1)-dimensions}},  {\em Phys.Lett.} {\bf B243} (1990) 378--382.

\bibitem{Campoleoni:2012hp}
A.~Campoleoni, S.~Fredenhagen, S.~Pfenninger, and S.~Theisen, {\it {Towards
  metric-like higher-spin gauge theories in three dimensions}},
  \href{http://xxx.lanl.gov/abs/1208.1851}{{\tt arXiv:1208.1851}}.

\bibitem{Banados:2012ue}
M.~Banados, R.~Canto, and S.~Theisen, {\it {The Action for higher spin black
  holes in three dimensions}},  {\em JHEP} {\bf 1207} (2012) 147,
  [\href{http://xxx.lanl.gov/abs/1204.5105}{{\tt arXiv:1204.5105}}].

\bibitem{Gaberdiel:2010pz}
M.~R. Gaberdiel and R.~Gopakumar, {\it {An AdS$_3$ Dual for Minimal Model
  CFTs}},  {\em Phys.Rev.} {\bf D83} (2011) 066007,
  [\href{http://xxx.lanl.gov/abs/1011.2986}{{\tt arXiv:1011.2986}}].

\bibitem{Gaberdiel:2011wb}
M.~R. Gaberdiel and T.~Hartman, {\it {Symmetries of Holographic Minimal
  Models}},  {\em JHEP} {\bf 1105} (2011) 031,
  [\href{http://xxx.lanl.gov/abs/1101.2910}{{\tt arXiv:1101.2910}}].

\bibitem{Gaberdiel:2011zw}
M.~R. Gaberdiel, R.~Gopakumar, T.~Hartman, and S.~Raju, {\it {Partition
  Functions of Holographic Minimal Models}},  {\em JHEP} {\bf 1108} (2011) 077,
  [\href{http://xxx.lanl.gov/abs/1106.1897}{{\tt arXiv:1106.1897}}].

\bibitem{Creutzig:2011fe}
T.~Creutzig, Y.~Hikida, and P.~B. Ronne, {\it {Higher spin AdS$_3$ supergravity
  and its dual CFT}},  {\em JHEP} {\bf 1202} (2012) 109,
  [\href{http://xxx.lanl.gov/abs/1111.2139}{{\tt arXiv:1111.2139}}].

\bibitem{Chang:2011vka}
C.-M. Chang and X.~Yin, {\it {Correlators in $W_N$ Minimal Model Revisited}},
  {\em JHEP} {\bf 1210} (2012) 050,
  [\href{http://xxx.lanl.gov/abs/1112.5459}{{\tt arXiv:1112.5459}}].

\bibitem{Candu:2012jq}
C.~Candu and M.~R. Gaberdiel, {\it {Supersymmetric holography on $AdS_3$}},
  \href{http://xxx.lanl.gov/abs/1203.1939}{{\tt arXiv:1203.1939}}.

\bibitem{Gaberdiel:2012ku}
M.~R. Gaberdiel and R.~Gopakumar, {\it {Triality in Minimal Model Holography}},
   {\em JHEP} {\bf 1207} (2012) 127,
  [\href{http://xxx.lanl.gov/abs/1205.2472}{{\tt arXiv:1205.2472}}].

\bibitem{Gaberdiel:2012uj}
M.~R. Gaberdiel and R.~Gopakumar, {\it {Minimal Model Holography}},
  \href{http://xxx.lanl.gov/abs/1207.6697}{{\tt arXiv:1207.6697}}.

\bibitem{Candu:2012tr}
C.~Candu and M.~R. Gaberdiel, {\it {Duality in N=2 Minimal Model Holography}},
  {\em JHEP} {\bf 1302} (2013) 070,
  [\href{http://xxx.lanl.gov/abs/1207.6646}{{\tt arXiv:1207.6646}}].

\bibitem{Creutzig:2012ar}
T.~Creutzig, Y.~Hikida, and P.~B. Ronne, {\it {N=1 supersymmetric higher spin
  holography on AdS$_3$}},  {\em JHEP} {\bf 1302} (2013) 019,
  [\href{http://xxx.lanl.gov/abs/1209.5404}{{\tt arXiv:1209.5404}}].

\bibitem{Candu:2012ne}
C.~Candu, M.~R. Gaberdiel, M.~Kelm, and C.~Vollenweider, {\it {Even spin
  minimal model holography}},  {\em JHEP} {\bf 1301} (2013) 185,
  [\href{http://xxx.lanl.gov/abs/1211.3113}{{\tt arXiv:1211.3113}}].

\bibitem{Peng:2012ae}
C.~Peng, {\it {Dualities from higher-spin supergravity}},  {\em JHEP} {\bf
  1303} (2013) 054, [\href{http://xxx.lanl.gov/abs/1211.6748}{{\tt
  arXiv:1211.6748}}].

\bibitem{Chang:2013izp}
C.-M. Chang and X.~Yin, {\it {A semi-local holographic minimal model}},
  \href{http://xxx.lanl.gov/abs/1302.4420}{{\tt arXiv:1302.4420}}.

\bibitem{Banados:1992wn}
M.~Banados, C.~Teitelboim, and J.~Zanelli, {\it {The Black hole in
  three-dimensional space-time}},  {\em Phys.Rev.Lett.} {\bf 69} (1992)
  1849--1851, [\href{http://xxx.lanl.gov/abs/hep-th/9204099}{{\tt
  hep-th/9204099}}].

\bibitem{Banados:1992gq}
M.~Banados, M.~Henneaux, C.~Teitelboim, and J.~Zanelli, {\it {Geometry of the
  (2+1) black hole}},  {\em Phys.Rev.} {\bf D48} (1993) 1506--1525,
  [\href{http://xxx.lanl.gov/abs/gr-qc/9302012}{{\tt gr-qc/9302012}}].

\bibitem{Gutperle:2011kf}
M.~Gutperle and P.~Kraus, {\it {Higher Spin Black Holes}},  {\em JHEP} {\bf
  1105} (2011) 022, [\href{http://xxx.lanl.gov/abs/1103.4304}{{\tt
  arXiv:1103.4304}}].

\bibitem{Ammon:2011nk}
M.~Ammon, M.~Gutperle, P.~Kraus, and E.~Perlmutter, {\it {Spacetime Geometry in
  Higher Spin Gravity}},  {\em JHEP} {\bf 1110} (2011) 053,
  [\href{http://xxx.lanl.gov/abs/1106.4788}{{\tt arXiv:1106.4788}}].

\bibitem{Kraus:2011ds}
P.~Kraus and E.~Perlmutter, {\it {Partition functions of higher spin black
  holes and their CFT duals}},  {\em JHEP} {\bf 1111} (2011) 061,
  [\href{http://xxx.lanl.gov/abs/1108.2567}{{\tt arXiv:1108.2567}}].

\bibitem{Ammon:2012wc}
M.~Ammon, M.~Gutperle, P.~Kraus, and E.~Perlmutter, {\it {Black holes in three
  dimensional higher spin gravity: A review}},  {\em J.Phys.} {\bf A46} (2013)
  214001, [\href{http://xxx.lanl.gov/abs/1208.5182}{{\tt arXiv:1208.5182}}].

\bibitem{Perez:2012cf}
A.~Perez, D.~Tempo, and R.~Troncoso, {\it {Higher spin gravity in 3D: black
  holes, global charges and thermodynamics}},
  \href{http://xxx.lanl.gov/abs/1207.2844}{{\tt arXiv:1207.2844}}.

\bibitem{David:2012iu}
J.~R. David, M.~Ferlaino, and S.~P. Kumar, {\it {Thermodynamics of higher spin
  black holes in 3D}},  {\em JHEP} {\bf 1211} (2012) 135,
  [\href{http://xxx.lanl.gov/abs/1210.0284}{{\tt arXiv:1210.0284}}].

\bibitem{Perez:2013xi}
A.~Perez, D.~Tempo, and R.~Troncoso, {\it {Higher spin black hole entropy in
  three dimensions}},  \href{http://xxx.lanl.gov/abs/1301.0847}{{\tt
  arXiv:1301.0847}}.

\bibitem{deBoer:2013gz}
J.~de~Boer and J.~I. Jottar, {\it {Thermodynamics of Higher Spin Black Holes in
  AdS$_{3}$}},  \href{http://xxx.lanl.gov/abs/1302.0816}{{\tt
  arXiv:1302.0816}}.

\bibitem{Kraus:2013esi}
P.~Kraus and T.~Ugajin, {\it {An Entropy Formula for Higher Spin Black Holes
  via Conical Singularities}},  {\em JHEP} {\bf 1305} (2013) 160,
  [\href{http://xxx.lanl.gov/abs/1302.1583}{{\tt arXiv:1302.1583}}].

\bibitem{Ammon:2013hba}
M.~Ammon, A.~Castro, and N.~Iqbal, {\it {Wilson Lines and Entanglement Entropy
  in Higher Spin Gravity}},  \href{http://xxx.lanl.gov/abs/1306.4338}{{\tt
  arXiv:1306.4338}}.

\bibitem{deBoer:2013vca}
J.~de~Boer and J.~I. Jottar, {\it {Entanglement Entropy and Higher Spin
  Holography in AdS$_3$}},  \href{http://xxx.lanl.gov/abs/1306.4347}{{\tt
  arXiv:1306.4347}}.

\bibitem{Castro:2012bc}
A.~Castro, E.~Hijano, and A.~Lepage-Jutier, {\it {Unitarity Bounds in AdS${}_3$
  Higher Spin Gravity}},  {\em JHEP} {\bf 1206} (2012) 001,
  [\href{http://xxx.lanl.gov/abs/1202.4467}{{\tt arXiv:1202.4467}}].

\bibitem{Castro:2011fm}
A.~Castro, E.~Hijano, A.~Lepage-Jutier, and A.~Maloney, {\it {Black Holes and
  Singularity Resolution in Higher Spin Gravity}},  {\em JHEP} {\bf 1201}
  (2012) 031, [\href{http://xxx.lanl.gov/abs/1110.4117}{{\tt
  arXiv:1110.4117}}].

\bibitem{Krishnan:2013cra}
C.~Krishnan and S.~Roy, {\it {Higher Spin Resolution of a Toy Big Bang}},
  \href{http://xxx.lanl.gov/abs/1305.1277}{{\tt arXiv:1305.1277}}.

\bibitem{Ouyang:2011fs}
P.~Ouyang, {\it {Toward Higher Spin dS3/CFT2}},
  \href{http://xxx.lanl.gov/abs/1111.0276}{{\tt arXiv:1111.0276}}.

\bibitem{deBuyl:2013ega}
S.~de~Buyl, S.~Detournay, G.~Giribet, and G.~S. Ng, {\it {Baby de Sitter Black
  Holes and dS$_3$/CFT$_2$}},  \href{http://xxx.lanl.gov/abs/1308.5569}{{\tt
  arXiv:1308.5569}}.

\bibitem{Cornalba:2002fi}
L.~Cornalba and M.~S. Costa, {\it {A New cosmological scenario in string
  theory}},  {\em Phys.Rev.} {\bf D66} (2002) 066001,
  [\href{http://xxx.lanl.gov/abs/hep-th/0203031}{{\tt hep-th/0203031}}].

\bibitem{Berkooz:2002je}
M.~Berkooz, B.~Craps, D.~Kutasov, and G.~Rajesh, {\it {Comments on cosmological
  singularities in string theory}},  {\em JHEP} {\bf 0303} (2003) 031,
  [\href{http://xxx.lanl.gov/abs/hep-th/0212215}{{\tt hep-th/0212215}}].

\bibitem{Johnson:2004zq}
C.~V. Johnson and H.~G. Svendsen, {\it {An Exact string theory model of closed
  time-like curves and cosmological singularities}},  {\em Phys.Rev.} {\bf D70}
  (2004) 126011, [\href{http://xxx.lanl.gov/abs/hep-th/0405141}{{\tt
  hep-th/0405141}}].

\bibitem{Balasubramanian:2002ry}
V.~Balasubramanian, S.~Hassan, E.~Keski-Vakkuri, and A.~Naqvi, {\it {A
  Space-time orbifold: A Toy model for a cosmological singularity}},  {\em
  Phys.Rev.} {\bf D67} (2003) 026003,
  [\href{http://xxx.lanl.gov/abs/hep-th/0202187}{{\tt hep-th/0202187}}].

\bibitem{Cornalba:2003kd}
L.~Cornalba and M.~S. Costa, {\it {Time dependent orbifolds and string
  cosmology}},  {\em Fortsch.Phys.} {\bf 52} (2004) 145--199,
  [\href{http://xxx.lanl.gov/abs/hep-th/0310099}{{\tt hep-th/0310099}}].

\bibitem{Craps:2007ch}
B.~Craps, T.~Hertog, and N.~Turok, {\it {On the Quantum Resolution of
  Cosmological Singularities using AdS/CFT}},  {\em Phys.Rev.} {\bf D86} (2012)
  043513, [\href{http://xxx.lanl.gov/abs/0712.4180}{{\tt arXiv:0712.4180}}].

\bibitem{Biswas:2011ar}
T.~Biswas, E.~Gerwick, T.~Koivisto, and A.~Mazumdar, {\it {Towards singularity
  and ghost free theories of gravity}},  {\em Phys.Rev.Lett.} {\bf 108} (2012)
  031101, [\href{http://xxx.lanl.gov/abs/1110.5249}{{\tt arXiv:1110.5249}}].

\bibitem{Bagchi:2012xr}
A.~Bagchi, S.~Detournay, R.~Fareghbal, and J.~Simon, {\it {Holography of 3d
  Flat Cosmological Horizons}},  {\em Phys.Rev.Lett.} {\bf 110} (2013) 141302,
  [\href{http://xxx.lanl.gov/abs/1208.4372}{{\tt arXiv:1208.4372}}].

\bibitem{Witten:1988hc}
E.~Witten, {\it {(2+1)-Dimensional Gravity as an Exactly Soluble System}},
  {\em Nucl.Phys.} {\bf B311} (1988) 46.

\bibitem{Achucarro:1987vz}
A.~Achucarro and P.~Townsend, {\it {A Chern-Simons Action for Three-Dimensional
  anti-De Sitter Supergravity Theories}},  {\em Phys.Lett.} {\bf B180} (1986)
  89.

\bibitem{DInverno}
R.~D'Inverno, {\it {Introducing Einstein's Relativity}}, . Oxford, UK: Univ.
  Pr. (1992) 383 p.

\bibitem{HawkingEllis}
S.~W. Hawking and G.~F. Ellis, {\it {The Large Scale Structure of
  Space-Times}}, . Cambridge, UK: Univ. Pr. (1975) 391 p.

\bibitem{Spradlin:2001pw}
M.~Spradlin, A.~Strominger, and A.~Volovich, {\it {Les Houches lectures on de
  Sitter space}},  \href{http://xxx.lanl.gov/abs/hep-th/0110007}{{\tt
  hep-th/0110007}}.

\bibitem{Tan:2011tj}
H.-S. Tan, {\it {Aspects of Three-dimensional Spin-4 Gravity}},  {\em JHEP}
  {\bf 1202} (2012) 035, [\href{http://xxx.lanl.gov/abs/1111.2834}{{\tt
  arXiv:1111.2834}}].

\bibitem{Ferlaino:2013vga}
M.~Ferlaino, T.~Hollowood, and S.~P. Kumar, {\it {Asymptotic symmetries and
  thermodynamics of higher spin black holes in AdS3}},
  \href{http://xxx.lanl.gov/abs/1305.2011}{{\tt arXiv:1305.2011}}.

\bibitem{Park:1998qk}
M.-I. Park, {\it {Statistical entropy of three-dimensional Kerr-de Sitter
  space}},  {\em Phys.Lett.} {\bf B440} (1998) 275--282,
  [\href{http://xxx.lanl.gov/abs/hep-th/9806119}{{\tt hep-th/9806119}}].

\bibitem{Balasubramanian:2001nb}
V.~Balasubramanian, J.~de~Boer, and D.~Minic, {\it {Mass, entropy and
  holography in asymptotically de Sitter spaces}},  {\em Phys.Rev.} {\bf D65}
  (2002) 123508, [\href{http://xxx.lanl.gov/abs/hep-th/0110108}{{\tt
  hep-th/0110108}}].

\bibitem{Krishnan:2013zya}
C.~Krishnan, A.~Raju, S.~Roy, and S.~Thakur, {\it {Higher Spin Cosmology}},
  \href{http://xxx.lanl.gov/abs/1308.6741}{{\tt arXiv:1308.6741}}.

\end{thebibliography}\endgroup

\end{document}